# Title: Psychological Effect of AI driven marketing tools for beauty/facial feature enhancement

Authors: Ayushi Agrawal[1], Aditya Kondai[1], Kavita Vemuri[1]

1. Cognitive Science Lab, International Institute of Information and Technology, Hyderabad.

**Extended abstract**

Introduction:

Facial beauty is often seen as universally understood (Rhodes,2006) and though it defies a single definition (Armstrong, 2004), we generally agree on what we find attractive (Webster and Driskell, 1983). The advent of AI-powered facial assessment tools has shifted how individuals evaluate their appearance, self-worth, and consumer choices. However, concerns arise about whether these tools, particularly in appearance-based marketing, contribute to increased self-objectification, lower self-esteem, and heightened focus on appearance enhancement(AE). Our study explores impact of such tools on a person's self-evaluation of appearance in relation to their self-objectification and self-esteem, their emotions, and assesses their perception of social emotion in others. It also explores gender differences in appearance-based social biases and judgments.

Methodology:

We conducted a longitudinal study with two samples using different versions of a facial assessment tool(Quoves — now Qoves — https://www.qoves.com). Sample-1(N=75; 29F, 46M; mean-age=22.9 years) used the older version(May 2021), which presented facial flaws overtly along with cosmetic correction recommendations. Sample-2(N=51; 25F, 26M; mean-age=19.9 years) used the newer version(May 2024), which described flaws as "concerns" and used a dermatological skin scale. Participants completed questionnaires assessing self-objectification (FQSO—Study1, SOS-T—Study2)and self-esteem (RSES). We also measured factors— Induced Emotion(IE), digital AE (DAE), physical AE and endorsement (PAEE), and perceived social emotion(PSE) in others—identified by factor analyses of the custom questions answered after receiving tool recommendations. Statistical analyses included McDonald's Omega Reliability test for consistency, Correlational analysis for associations, Levene's test for variance, and Mann-Whitney U test to evaluate differences across tool versions and gender.

Results and discussion:

The FQSO, SOS-T, and RSES responses showed internal consistency. Variance was equal across parameters, except for FQSO/SOS-T between studies and PAEE and FQSO/SOS-T across gender. In Study 1, FQSO positively correlated with DAE and PAEE, but negatively with RSES, indicating that lower self-esteem was linked to higher self-objectification. In Study 2, DAE was positively correlated with PAEE and SOS-T, while PSE and RSES were negatively correlated. The significant difference in DAE and PAEE association between studies (Spearman's rho:Study1=0.626,Study2=0.285,z=2.3,p=0.0178) suggests a diminished relationship over time. The increase in DAE from Study1(mean=2.844) to Study2(mean=3.007) indicates a growing tendency to use digital filters, possibly due to social pressure. DAE's significant correlation with self-objectification scales across both studies show that individuals with higher self-objectifying attitudes are more inclined to digitally enhance their appearance. Mann-Whitney showed a significant difference in IE between studies (m1=5.083,m2=4.510,U=1466.5,p=0.013), indicating that despite subtler flaw presentation in Study 2, participants felt worse. This suggests that subtle feedback may not always reduce negative emotions and could exacerbate insecurity, especially in individuals with higher self-objectification tendencies. Gender differences were significant in DAE(p=0.025) and PSE(p<0.001), with females reporting higher DAE(mean=3.611) and lower PSE (mean=5.710) than males.

Conclusion and Limitations:

This study highlights how AI-powered facial assessment tools may exacerbate self-objectification and reinforce appearance-based social judgments independent of their presentation of analysis, especially among vulnerable groups. Gender differences reveal distinct interactions with tools: Females perceived induced insecurity in others less than males and were more likely to alter their digital appearance, suggesting they may be more accepting of appearance-based social judgments and biases. Future research should explore how human ideologies embedded in the training data of these tools shape their evaluative outputs and influence user attitudes and decisions. It should also include more diverse populations to address the limitations of demographic homogeneity, investigate the long-term psychological effects of repeated tool use, and work toward standardizing self-objectification measures to enable better cross-study comparability.

Abbreviations:
Female Questionnaire of Trait Self-Objectification—FQSO, Self-Objectification Scale-Trait—SOS-T, Rosenberg Self-Esteem Scale—RSES.